%% file: main.tex
\documentclass[floatfix,lengthcheck,showpacs,amssymb,amsmath,amsfonts,twocolumn,nofootinbib,longbibliography]{revtex4-1}
\usepackage{amsfonts,amsmath,units,wasysym,epsfig,graphicx,verbatim,color,subfigure,graphicx}
\usepackage{amsmath}
\usepackage{latexsym}
\usepackage{amssymb}
\usepackage{amsfonts}
\usepackage{mathtools}
\usepackage{bm}
\usepackage{color}
\usepackage{float}
\usepackage{tikz}
\usepackage{adjustbox}
\usepackage{array}
\usepackage{soul}
\usepackage{appendix}
\usepackage{physics}
\usepackage{braket}
\usepackage{xcolor}
\usepackage[none]{hyphenat}
\usepackage{longtable}
\usepackage{lineno}
\usepackage{placeins}

\newcommand{\approximant}[1]{{\fontfamily{qcr}\selectfont{#1} }}

\begin{document}

\author{Rahul Dhurkunde} 
\affiliation{Institute of Cosmology and Gravitation, University of Portsmouth, Portsmouth, PO1 3FX, U.K.}
\author{Ian Harry}
\affiliation{Institute of Cosmology and Gravitation, University of Portsmouth, Portsmouth, PO1 3FX, U.K.}

\date{\today}
\title{Search for Precessing Binary Black Holes in Advanced LIGO’s Third Observing Run using Harmonic Decomposition}

\begin{abstract}
Binary black holes (BBHs) exhibiting spin-induced orbital precession offer unique insight into compact-binary formation channels, cosmology, and tests of general relativity. We conduct a dedicated search for precessing BBHs with unequal masses in Advanced LIGO’s third observing run (O3) using the harmonic decomposition method of precessing waveforms. We introduce a novel scheme to reduce the number of filters in a harmonic search. With our new approach, our template bank requires $5\times$ fewer filters compared to another proposed precessing search in the same region. We do not find any new significant events. Our new search method achieves up to $\sim 28\%$ improvement in sensitivity and up to $5\times$ lower computational costs compared to existing precessing search pipelines. Our method enables scalable, sensitive searches for precessing BBHs in future gravitational-wave observing runs. 
\end{abstract}
 \maketitle

\section{Introduction}

The discovery of gravitational waves (GWs) has improved our understanding of compact binary systems -- binaries involving black holes and/or neutron stars \cite{LIGOScientific:2025pvj}. When the spins of these black holes are misaligned with the orbital angular momentum, the orbital plane undergoes precession \cite{Apostolatos:1994mx}. This spin-orbit coupling effect leads to characteristic amplitude and phase modulations in the emitted gravitational waves, which are strongest for binaries with large in-plane spin components, asymmetric masses, or edge-on orientations \cite{CalderonBustillo:2016rlt, Harry:2016ijz, Pratten:2020igi, Green:2020ptm}. Such configurations are expected in a variety of astrophysical scenarios.

Precessing binaries can form through multiple evolutionary channels \cite{Mandel:2018hfr, Mapelli:2021taw, Gerosa:2021mno}. In isolated binary evolution, supernova kicks imparted during the formation of compact objects can misalign the spins from the orbital angular momentum \cite{Kalogera:1999tq, Gerosa:2018wbw}. Alternatively, in dynamically formed systems—such as those originating in globular clusters, nuclear star clusters, or hierarchical triples—the orientations of spins are expected to be nearly isotropic \cite{Belczynski:2014iua,Rodriguez:2016vmx,Mandel:2018hfr}. 

The detection and precise characterization of precessing binaries have significant implications for several domains. From an astrophysical perspective, clear measurement of precession can offer insights into the formation environments of compact binaries \cite{KAGRA:2021duu, LIGOScientific:2025pvj}. Observational evidence of spin-induced precession is a more likely signature of a dynamical formation over an isolated channel, and may help estimate the relative rates of the various formation channels. For cosmology using GWs, precession can help break the inclination-distance degeneracy \cite{Raymond:2008im}, resulting in better estimates of the redshifts for these sources, which in turn can give better estimates of the Hubble constant via Dark vs. Bright siren cosmological methods \cite{Yun:2023ygz}. Furthermore, precessing binaries can provide cleaner tests of general relativity in the strong-field regime by breaking mode degeneracies and enhancing the spectral sensitivity of ringdown tests to deviations from the Kerr metric \cite{Zhu:2023fnf}. Precessing signals can also provide tests of the large-scale symmetries of the Universe \cite{CalderonBustillo:2024akj}.

As of the latest release, the Gravitational-Wave Transient Catalog 4 (GWTC-4) \cite{LIGOScientific:2025slb} contains nearly two hundred well-characterized binary mergers detected by the Advanced LIGO \cite{LIGOScientific:2014pky} and Advanced Virgo \cite{VIRGO:2014yos} detectors. The majority of these events are consistent with aligned-spin systems \cite{LIGOScientific:2025pvj}. A population-level analysis comparing observations from the catalogs GWTC-3 \cite{KAGRA:2021vkt} to GWTC-4 reveals an increasing preference for systems with $\chi_P > 0$ (see Figure 8 of \cite{LIGOScientific:2025pvj}), where $\chi_P$ is the effective precession parameter introduced in \cite{Schmidt:2014iyl}. A few events show potential evidence for orbital precession, with GW200129\_065458 to be the strongest so far: the event has been measured to have a precessing frequency (the rate at which the orbital plane rotates) of at least 1 Hz when the signal enters the LIGO sensitive band \cite{Hannam:2021pit}. The scarcity of confidently identified precessing signals may point to intrinsic rarity of the precessing binary population \cite{Hoy:2024wkc} or reflect the limitations of current detection pipelines which omit spin-precession effects. 

Modeled gravitational-wave searches typically use a bank of filters (templates) to perform matched filtering of the interferometric data \cite{DalCanton:2014hxh, Usman:2015kfa, Messick:2016aqy, Nitz:2017svb, Allene:2025saz}. Standard searches use template banks restricted to aligned-spin configurations due to computational efficiency and a simpler matched-filter statistic. For an aligned-spin search, the effects of varying the unknown extrinsic parameters of a binary -- such as sky-location and inclination angle of the binary plane can be incorporated as a fiducial amplitude and overall phase to the gravitational-wave signal. In such cases, the matched filter can be conveniently performed with templates that depend only on four parameters -- component masses and spins restricted parallel to the orbital angular momentum. The same does not hold true for precessing systems, as the signal seen by a detector is phase and/or amplitude modulated by varying inclination angle, sky-location or due to time-varying component spins. Aligned-spin template banks have poor ability to capture generically precessing signals due to omission of precessional effects \cite{CalderonBustillo:2016rlt, Harry:2016ijz, Chandra:2020ccy, Dhurkunde:2022aek, Schmidt:2024jbp}. Specifically for highly precessing BBH systems, aligned-spin banks can lose up to $25\%$ of signal-to-noise ratio (SNR) \cite{Schmidt:2024jbp}. Although unmodeled pipelines and semi-analytical strategies exist, they struggle to match the sensitivity of matched-filtering techniques for low-mass systems \cite{Nitz:2021zwj, Olsen:2022pin, LIGOScientific:2025yae}.

Several methods have been proposed to detect precessing compact binary mergers in the past, but most of them have struggled with balancing sensitivity with computational cost \cite{Grandclement:2002dv, Pan:2003qt, Buonanno:2004yd, Buonanno:2005pt, VanDenBroeck:2009gd, Harry:2016ijz}. Precession effects are represented by modulating the phase and amplitude of non-precessing signals: these modulations require additional parameters. In \cite{Grandclement:2002dv}, these additional parameters were added to the template bank but the templates failed to yield adequate matches. Later approaches allowing analytical maximization over the free parameters led to unphysical solutions and higher false-alarm rates \cite{VanDenBroeck:2009gd}. Fully physical template banks covering all parameters were also attempted but proved computationally infeasible, requiring tens of millions of templates \cite{Harry:2016ijz}. Despite the challenges, there have been two precessing searches performed on the O3 data from Advanced LIGO detectors: one for binary black holes (BBHs) \cite{Schmidt:2024hac}, and for neutron star-black hole binaries \cite{Harry:2025eqk}. No new candidates were found w.r.t the existing aligned-spin searches \cite{LIGOScientific:2025slb} in any of the precessing searches. 

The decomposition of any precessing signal into multiple harmonics offers a promising path to perform precessing searches \cite{Fairhurst:2019vut}. Related formalisms for single-spin binaries were introduced earlier in \cite{Lundgren:2013jla, OShaughnessy:2015xjs}. As recently demonstrated in \cite{Harry:2025eqk}, the harmonic precessing search for NSBH systems requires $3\times$ fewer templates than brute-force approaches, and is up to $100\%$ more sensitive than an equivalent aligned-spin search. In this work, we use an improved implementation of the same harmonic decomposition technique to perform a full-scale search for highly precessing BBH systems. We introduce a new technique that substantially reduces the number of filters required for the harmonic search. With this method, we achieve a fivefold reduction in filters compared to the previous search for highly precessing BBHs \cite{Schmidt:2024hac}, while simultaneously achieving significantly improved fitting factors across the template bank.

\begin{figure}[]
    \centering
    \includegraphics[width=\linewidth]{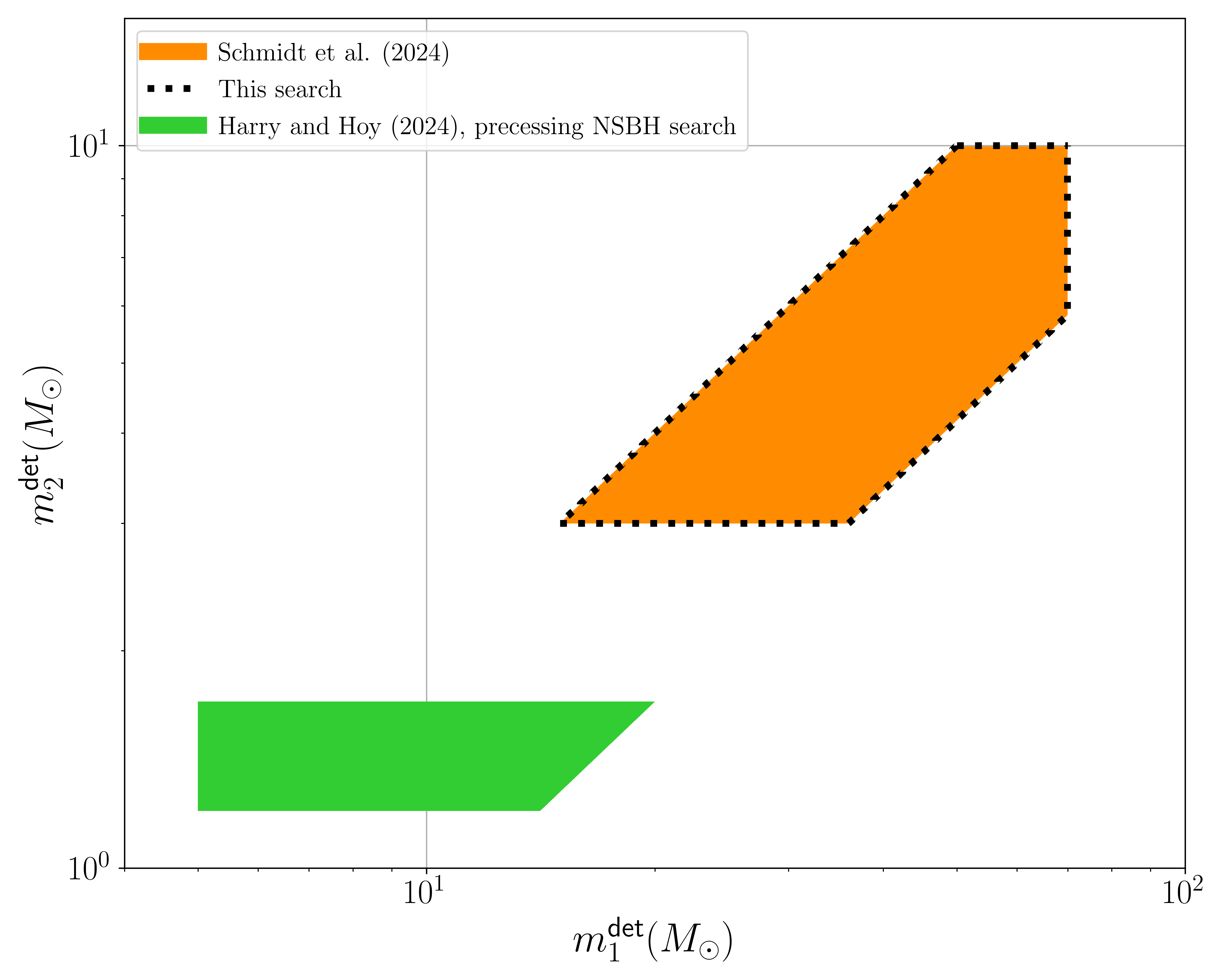}
    \caption{Search region of the previous precessing searches in the O3 data of the Advanced LIGO detectors for compact binary mergers to-date. Shown is the search region as the two dimensional space of detector frame component masses $(m_{1}^{\text{det}}, m_{2}^{\text{det}})$. The green area denotes the parameter space explored in the recent NSBH precessing search that employed the same harmonic decomposition technique \cite{Fairhurst:2019vut, McIsaac:2023ijd}. The orange area corresponds to the region used in the highly precessing BBH search of \cite{Schmidt:2024hac}. We have chosen the exact same parameter region as the previous precessing BBH search (marked as black dotted lines).}
     \label{fig:past-searches}    
\end{figure}

In this work we search using publicly available data from the Advanced LIGO detectors during their third observing run \cite{KAGRA:2023pio}. We choose the same parameter space as the most recent search for precessing BBHs \cite{Schmidt:2024hac} Our search region covers as shown in Figure \ref{fig:past-searches} : detector frame component masses $m_1^{\text{det}} \in [15, 70] M_{\odot}$, $m_2^{\text{det}} \in [3, 10] M_{\odot}$ with a cutoff on mass-ratio $q = m_1^{\text{det}}/m_2^{\text{det}} \in [5, 12]$. We did not find any new significant events and recover $29$ events that were previously reported in the latest GWTC-4 catalog \cite{LIGOScientific:2025pvj}. Our precessing search combined with an equivalent aligned-spin search in the same mass region, achieves up to $28\%$ higher sensitivity compared to an aligned-spin only search, while using $5\times$ fewer filters than a previous precessing search in the same region. 


\section{Search for orbital precession using harmonics}


The harmonic–decomposition formalism for detecting single-spin precessing systems was introduced in \cite{Lundgren:2013jla, OShaughnessy:2015xjs} and for fully-precessing systems in \cite{Fairhurst:2019vut}, the latter was implemented as a complete end-to-end search framework in \cite{McIsaac:2023ijd}, and subsequently applied to advanced LIGO data to search for precessing NSBH binaries in \cite{Harry:2025eqk}. In this work, we primarily build on the implementation of \cite{McIsaac:2023ijd}, supplemented with new enhancements that further improve the efficiency of the harmonic search. Sections \ref{Sec:harmonic-decomposition}–\ref{Sec:search-description} review the methodology for conducting a harmonic-based search, and Section \ref{Sec:our-improvements} details the improvements introduced in this study.



\subsection{Harmonic decomposition of precessing waveforms}\label{Sec:harmonic-decomposition}

Consider a precessing binary system in a quasi-circular orbit with an orbital angular momentum $\textbf{L}$ and spin angular momentum vectors for each component $\textbf{S}_1, \textbf{S}_2$. The total angular momentum is given by $\textbf{J} = \textbf{L} + \textbf{S}_1 + \textbf{S}_2$. The gravitational waveform emitted by the binary as seen by the detector is expressed approximately as \cite{Buonanno:2002fy}
\begin{align}
    h(t) =  &-\Big(\dfrac{d_0}{d_L}\Big) A_0(t) \times \\
     & \Re\Big[ e^{2i\Phi_S(t)} \Big( F_+(C_+ -iS_+) + F_{\times}(C_{\times} - iS_{\times}) \Big) \Big], \nonumber
    \label{Eq:first-GW-eq}    
\end{align}
where $A_0(t)$ is the time-varying amplitude which depends upon the component masses and spins of the binary, and we have separated the luminosity distance $d_L$ dependence as the waveform defined at a fiducial distance $d_0$. $F_+$ and $F_{\times}$ are the detector response (relative to \textbf{J}-frame) functions, and $C_{+,\times}, S_{+,\times}$ are the time-varying response to the gravitational wave. 

The phase $\Phi_S(t)$ is defined in the frame aligned with \textbf{J}, and its evolution is related to the orbital phase $\phi_{orb}$ as
\begin{align}
    \Phi_s(t) = \phi_{orb}(t) - \epsilon(t),
\end{align}
where the binary precesses with rate $\dot{\epsilon}$ given as
\begin{align}
    \dot{\epsilon}(t) = \dot{\alpha}\cos\beta(t),
\end{align}
and $\beta$ is the opening angle between $\textbf{J}$ and \textbf{L} and $\dot{\alpha}(t)$ is the precession frequency. Using the parameter 
\begin{align}
    b = \tan(\beta/2),
\end{align}
one can write the responses $C_{+,\times}, S_{+,\times}$ in terms of ($\theta_{\text{JN}}, \beta, \alpha$) where $\theta_{\text{JN}}$ is the angle between $\textbf{J}$ and the line-of-sight to the observer. Following \cite{Fairhurst:2019vut}, rewriting the equation (\ref{Eq:first-GW-eq}) as
\begin{align}
    h(t) = \Re \Bigg[ \Bigg( & \dfrac{A_0(t)e^{2i(\Phi_s+\alpha)}}{(1+b^2)^2}\Bigg) \\
    &\sum_{k=0}^{4} (be^{-i\alpha})^k (F_{+}\mathcal{A}_k^{+} - iF_{\times}\mathcal{A}_{k}^{\times})\Bigg] \nonumber,
\end{align}
where the amplitudes $\mathcal{A}_k^{+,\times}$ depend on $\theta_{\text{JN}}$ and $d_L$ and are explicitly given in \cite{Fairhurst:2019vut}. Factorizing out the dependence of orbital $\phi_{orb}$ and precession phase $\Phi_s$, the binary's phase evolution is written as \cite{Fairhurst:2019vut}
\begin{align}
    \Phi(t) = \Phi_s(t) - \Phi_0 + \alpha(t) - \alpha_0.
\end{align}
Following \cite{McIsaac:2023ijd}, we can then write the observed signal as
\begin{align}
    h(t) = \sum_{k=0}^{4} A_k h_k(t)e^{i\phi_k(t)},
    \label{eq:five-harm-decomp}
\end{align}
where
\begin{align}
    h_k(t) &= \dfrac{A_0(t)b^k(t)}{(1+b^2(t))^2}e^{i\big[2\Phi(t)-k(\alpha(t)-\alpha_0)]}, 
    \label{eq:harmonic-components} \\ 
    A_k &= ((F_+\mathcal{A}_k^{+})^2 + (F_{\times}\mathcal{A}_k^{\times})^2)^{1/2},\label{Eq:harm-amplitude}
\end{align}
and 
\begin{align}
    \phi_k = 2\Phi_0 + (2-k)\alpha_0 - \tan^{-1}\Bigg(\dfrac{F_{\times}\mathcal{A}_k^{\times}}{F_+\mathcal{A}_k^{+}} \Bigg)
    \label{Eq:harmonic-phase}
 \end{align}
From the equations (\ref{eq:five-harm-decomp}), (\ref{Eq:harm-amplitude}) and (\ref{Eq:harmonic-phase}), we see that any variation in the extrinsic parameters changes only the overall amplitudes and phases of each harmonic which do not vary over time. We note that, for a generic precessing binary, $\beta$ can range between $[0, \pi)$, and therefore, $b \in [0, \infty)$. For aligned / anti-aligned systems $b$ is either close to zero or is a large positive number, for maximally precessing configuration $b \sim 1$. From equation (\ref{eq:harmonic-components}), we see that the harmonic amplitudes roughly scale proportionally to $b^k$. Similar to the time-domain harmonics, the Fourier domain ones can be obtained by using the stationary phase approximation, and the precessing waveform can be written as
\begin{align}
    \tilde{h}(f) = \Re\sum_{k=0}^4 A_{k}\tilde{h}_k(f)e^{i\phi_k},
    \label{Eq:fourier-harmonic-components}
\end{align}
where $\tilde{h}(f)$ is the Fourier transform of $h(t)$. 


\subsection{Filtering with orthonormalized harmonics}\label{Sec:filtering-with-harmonics}

Modeled searches employ the matched filtering technique to estimate the likelihood of data $s$ containing a modeled gravitational signal $h$ \cite{Allen:2005fk,Babak:2012zx}. Matched filter is an optimal detection statistic under the assumption that data contains \textit{Gaussian} and \textit{stationary} noise. The matched filter is performed in the Fourier domain, by correlating the Fourier-transformed data $\tilde{s}(f)$ and a template $\tilde{h}(f)$ of the anticipated signal, weighted by the noise power spectral density (PSD) $S_n(f)$. The complex matched filter statistic is given by
\begin{align}
    \langle h|s\rangle &= 4 \int \dfrac{\tilde{h}^*(f) \tilde{s}(f)}{S_n(f)}df, \\
    (h|s) &= \Re[\langle h | s \rangle].
\end{align}
The signal-to-noise ratio (SNR) is defined as
\begin{align}
    \rho^2 = \dfrac{(h|s)^2}{(h|h)}.
\end{align}

A binary system in a quasi-circular orbit is characterized by 15 parameters: component masses $(m_1,m_2)$, component spin vectors $\vec{s}_1, \vec{s}_2$, inclination angle $i$, sky-location angles $(\theta, \delta)$, polarization $\psi$, luminosity distance to the source $d_L$, the orbital phase $\phi_c$ and time $t_c$ at binary coalescence. A priori, the signal parameters are unknown, so the matched filter statistic is maximized over a wide range of possibilities to maximize the likelihood of finding a real gravitational-wave signal. Therefore, a discrete set of filters (templates) are used to search over a given region of the parameter space. For aligned-spin signals, varying extrinsic parameters such as inclination and sky-location change only the amplitude or an overall phase to the signal. In such cases, all the extrinsic parameters except $(m_1,m_2,s_{1z}, s_{2z})$ are analytically maximized by taking the absolute of the complex matched filter output
\begin{align}
    \max(\rho^2) = |\langle h|s \rangle|^2.
\end{align}
The remaining parameters are included in the template bank. 

To perform matched filtering for precessing signals, we use the Fourier transformed harmonic components described in (\ref{Eq:fourier-harmonic-components}). If the five harmonics were independent, each could be matched-filtered separately while maximizing over the respective fiducial phase $\phi_k$ and amplitude $A_k$ parameters. However, their non-orthogonality introduces covariance among the matched-filter outputs, complicating the maximization of the SNR. To simplify this, we first orthonormalize the harmonics using Gram-Schmidt orthogonalization, according to the following equations:
\begin{align}
    h_k^{\perp} &= h_{k} - \sum_{l = 0}^{k-1}(h_{l}|h_k)h_{k}, \\ 
    \hat{h}_k &= \dfrac{h_k^{\perp}}{(h_k^{\perp}|h_k^{\perp})}.
\end{align}

\begin{figure*}[]
    \centering
    \includegraphics[width=\linewidth]{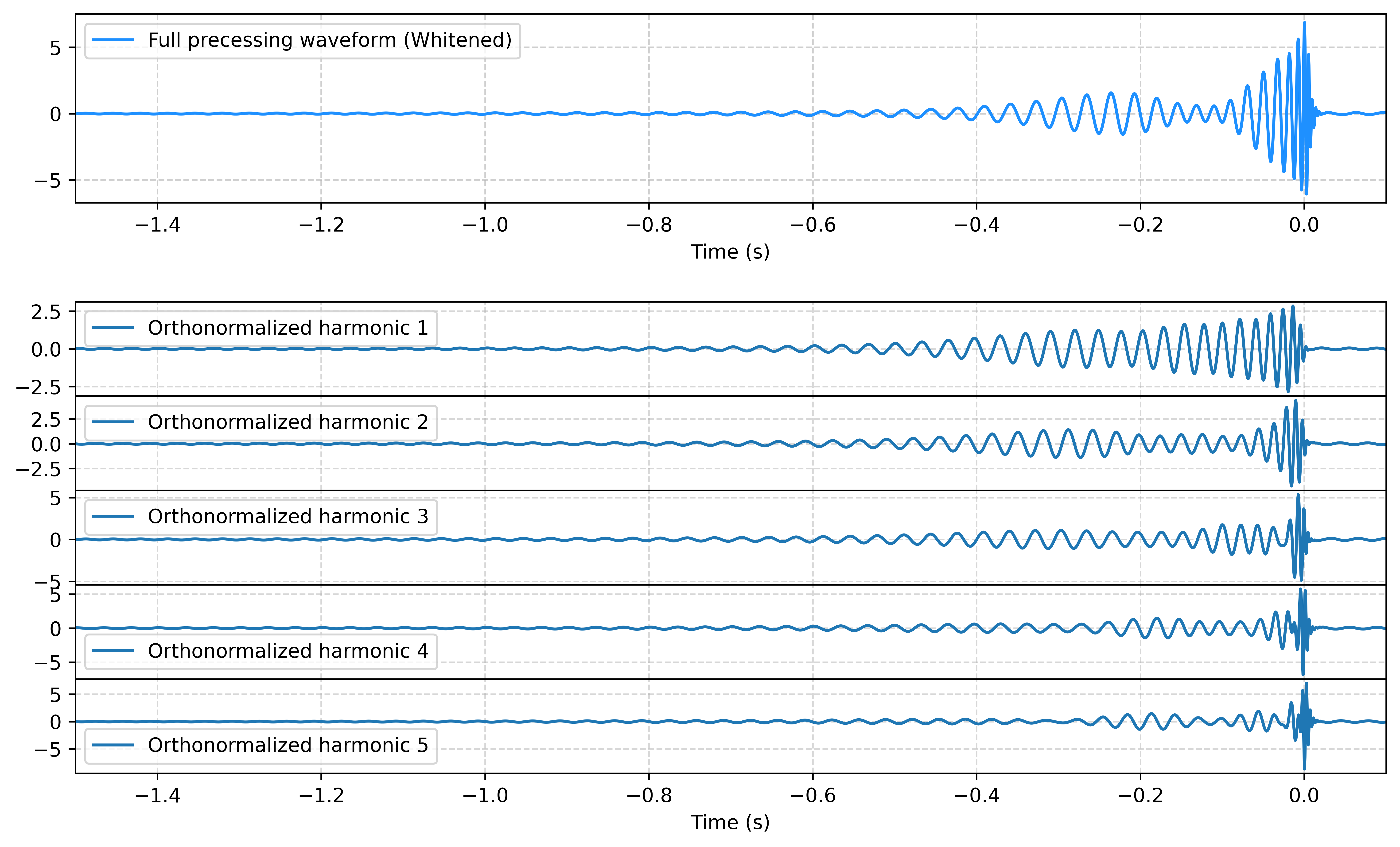}
    \caption{Whitened precessing waveform and the orthonormalized harmonic decomposition of the same waveform in the time-domain. On the top is the full whitened precessing waveform for a binary with $(m_1^{det}, m_2^{det} ) = (57.5, 5.7) M_{\odot}$, inclination $\sim 74^{o}$ and $b = 0.93$ generated using \approximant{IMRPhenomXP}. The second to sixth rows correspond to the orthonormalized harmonics, with the first being the dominant harmonic, followed by second then third and so on.}
     \label{fig:five-harmonics}    
\end{figure*}

An example of orthonormalized harmonics is shown in the Figure \ref{fig:five-harmonics}. We use these new orthonormalized harmonics, and filter each of them separately for the corresponding templates. The SNR for each harmonic is obtained by independently maximizing over the unknown amplitudes $A_k$ and phases $\phi_k$ and then combining in quadrature to get the total SNR for the given template
\begin{align}
    \rho_h^2 = \sum_{k = 0}^4 \max_{A_k, \phi_k}[\rho_k]^2,
\end{align}
where $\rho_k=(h_k^{\perp}|s)$ is the matched-filter output by filtering using the $k^{th}$ orthogonalized harmonic. In the next section we describe the procedure for constructing a harmonic template bank.

\subsection{Constructing the template bank}\label{Sec:TB-construction}

A bank of templates is used to search over the intrinsic parameters of the expected signal. Several sampling strategies exist in the literature to ensure adequate coverage of the parameter space \cite{Babak:2008rb, Harry:2009ea}; only a small fraction of injections within the bank’s domain should have a best match below the prescribed minimum match. The best match achieved between a signal and the templates in the bank is known as the fitting factor, which corresponds to the fraction of a signal $g$’s SNR that is recovered by a given template $h$:

\begin{align}
    \text{FF}(g) = \max_{i \in \text{bank}} [m(h_i, g)].
\end{align}

The stochastic sampling process initializes with an empty bank. A first template is selected randomly from the parameter space, after which successive templates are iteratively added after checking the match of a proposed template $h_{\text{prop}}$ against the harmonic decomposition of the existing templates $h_{i}$ in the bank. Match is maximized over the extrinsic parameters via maximizing over the fiducial parameters $A_k$ and $\phi_k$. We use a random sky-location and orientation of the proposed template to compute the match with existing templates in the bank using all five harmonics. If the fitting factor is still less than the mismatch, then $h_{\text{prop}}$ is added to the bank. 

In this search we consider only the dominant mode of gravitational emission. We use the \approximant{IMRPhenomXP} approximant to model precessing signals \cite{Pratten:2020fqn,Garcia-Quiros:2020qpx}. While computing the matches, we generate the templates with a lower frequency cutoff of 15Hz and use the O3 optimistic PSD \cite{KAGRA:2013rdx}. We use the sbank library to generate our bank with a minimum match = 0.97.  Our template bank spans 8 dimensions: detector frame component masses $(m_1^{\text{det}}, m_2^{\text{det}})$ and generic component spins $(s_{1x}, s_{1y}, s_{1z}, s_{2x}, s_{2y}, s_{2z})$. In our bank, we also store the fixed sky-location $(\theta, \delta)$, inclination $i$, polarization $\psi$ and coalescence phase $\phi_c$ used while computing the matches for generating the bank. The parameter ranges used for our bank are described in Table \ref{Table:TB-params}.

\begin{table}[]
    \centering
    Precessing bank
    \begin{tabular}{|c|c|}
    \hline 
        Parameter & Range \\
        \hline
        \hline
        $m_1$ & [15, 70]$M_{\odot}$\\
        \hline
        $m_2$ & [3, 10]$M_{\odot}$\\
        \hline
        $q$ & [5, 12] \\
        \hline
        $|s_1|$ & [0.5, 0.9] \\
        \hline
        $|s_2|$ & [0.0, 0.99] \\
        \hline
        $\phi_c$ & $[0, 2\pi]$  \\
        \hline
        $i$ & [$0, \pi$] \\
        \hline
        $(\alpha, \delta)$ & ($[0, 2\pi]$, $[0, \pi]$)\\
        \hline
        $\psi$  & $[0, 2\pi]$ \\
        \hline
    \end{tabular}\\
Aligned-spin bank
    \begin{tabular}{|c|c|}
    \hline 
        Parameter & Range \\
        \hline
        \hline
        $m_1$ & [15, 70]$M_{\odot}$\\
        \hline
        $m_2$ & [3, 10]$M_{\odot}$\\
        \hline
        $q$ & [5, 12] \\
        \hline
        $|s_{1z}|$ & [0.0, 0.99] \\
        \hline
        $|s_{2z}|$ & [0.0, 0.99] \\
        \hline
    \end{tabular}
    \caption{Description of the template banks used in this work -- our precessing bank and an equivalent aligned-spin bank. We show the parameters that are used to create the discrete set of templates and their respective ranges. Both template banks are created with a stochastic sampling method with a minimum match requirement of 0.97. Our precessing bank contains 8 parameters: component masses and generic spins. We assume random orientation and sky-location angles for a proposal template when comparing matches with existing templates during bank generation. Our bank has in total 211,730 templates. The equivalent aligned-spin bank has only 4 parameters with spins restricted to only in the $z$ direction and contains 33099 templates. }
    \label{Table:TB-params}
\end{table}

Our precessing template bank contains 211,730 templates—an order of magnitude smaller than the bank used in the previous precessing BBH search \cite{Schmidt:2024hac}, which comprised approximately $2.3 \times 10^6$ templates. The bank in \cite{Schmidt:2024hac} was constructed with a mismatch of 0.9; applying the same method with a mismatch of 0.97 would produce more than 30 million templates.

For comparison, we also construct an aligned-spin template bank by sampling over the same parameter region but restricting the spins to be aligned with the orbital angular momentum, with spin magnitudes constrained to $(|s_{1z}|, |s_{2z}|) \in [0.0, 0.99)$. The templates are generated using \approximant{IMRPhenomXAS} \cite{Pratten:2020fqn}, employing the same lower-frequency cutoff and PSD for all match calculations. This aligned-spin bank spans the four parameters $(m_1^{\text{det}}, m_2^{\text{det}}, s_{1z}, s_{2z})$ and consists of 33,099 templates.

\subsection{Reducing the number of filters}\label{Sec:reducing-the-filters}
Using all five harmonics for each template is computationally expensive, and is not necessary as not every template requires all five harmonics: templates which are aligned-spin like can be accurately represented by the dominant harmonic only, whereas precessing templates (significant in-plane spin components) would require more than just the dominant harmonic. Therefore, we pre-compute the number of harmonics required by each template. We estimate this by computing an \textit{effective match} $m_{\text{eff}}$ of each template $h_i$, which is an averaged match over a sufficient $n$ random sky-location and orientation (denoted by $\Omega$) for the same system i.e. keeping the other parameters fixed. The effective match is computed as
\begin{align}
    m_{\text{eff}}(h_i) = \bigg[\dfrac{m(h_i, h_i(\Omega))^3 \rho^3_{\text{opt}}}{\sum_{i=0}^n \rho^3_{\text{opt}}(h_i)}\bigg]^{1/3},
\end{align}
where $\rho_{\text{opt}}$ is the optimal SNR of the given template with itself or in other words the complete set of the orthonormalized harmonics
\begin{align}
    \rho_{\text{opt}} = \bigg[ \sum_{k=0}^{4} |(\hat{h}_k^{\perp}|h)|^2 \bigg]^{1/2}.
    \label{Eq:orthogonalization}
\end{align}

We calculate the effective match by successively increasing the number of harmonics until it reaches at least the minimum match  ($m_{\text{eff}} \geq 0.97$). We store the number of components for each template in our bank.

\subsection{Search Pipeline and Coincidence Analysis}\label{Sec:search-description}

We use the harmonic search implementation \cite{McIsaac:2023ijd} in the open-source software PyCBC \cite{alex_nitz_2024_10473621} to perform our precessing search. We restrict the number of harmonics to three to have a good compromise between the increasing computational costs of matched filtering and the ability to capture precessional effects from the data. We matched filter single detector data using the number of harmonics associated with each template to get SNR time-series for each harmonic. Then we add each harmonic SNR series in quadrature to the complete SNR time-series for the corresponding template.

To mitigate the effect of non-Gaussian noise artifacts, we perform two non-Gaussian vetoing tests: a modified version of the $\chi^2$ test (as implemented in \cite{McIsaac:2023ijd}) and an optimized version of the sine-Gaussian $\chi^2$ test \cite{Nitz:2017lco} (discussed in the next section). The standard $\chi^2$ test checks for differences in the expected and computed power in different frequency bands. For our precessing search, we compute the expected power for a reconstructed signal from a trigger. The reconstructed signal is a combination of the harmonics (up to three) weighted by their respective SNRs. Furthermore, we use the PSD variation statistics to incorporate short-term PSD variability \cite{Mozzon:2020gwa}. The SNR, $\chi^2$, sine-Gaussian and PSD variation statistics are combined together to rank the triggers from a single detector.

We then perform a coincident search between data from the two detectors -- LIGO Hanford and LIGO Livingston and use the multi-detector ranking statistic implemented in \cite{McIsaac:2023ijd}. The ranking statistic is constructed using the allowed coincidence window between detectors, the expected noise-rate density, and the likelihoods of extrinsic parameters under the signal and noise hypotheses. These likelihoods are derived by comparing the amplitudes and phases of a single harmonic across detectors, ensuring both share the same $b_k$ value, thereby removing dependence on intrinsic template parameters. The harmonic with the highest signal-to-noise ratio serves as the reference for evaluating amplitude ratio, phase difference, and time difference relative to the second detector. 


For our aligned-spin search, we use the same search configuration and ranking statistic as used for the PyCBC analysis of the O3 data for producing the GWTC-3 catalog \cite{KAGRA:2021vkt}. The aligned-spin search ranking statistic is further enhanced by incorporating an additional KDE-based statistic, which accurately models the probability distributions of binary source parameters across the template bank \cite{Kumar:2024bfe}.

\subsection{Improvements to the Harmonic Search}\label{Sec:our-improvements}

The previous scheme to compute the harmonics works naturally for templates that have $b < 1$: in the \textit{original} harmonic order with $h_0$ being the dominant harmonic. However, this is not true for $b\geq1$: since the harmonic amplitudes scale roughly with $b^k$ as the harmonic order is \textit{reversed}, the fifth harmonic will have the dominant contribution followed by the fourth, third and so on. The effective match with the original harmonics for $b \geq 1$ templates is greater than the minimum match only when the later harmonics are included. The original harmonic order may lead to a larger number of harmonics than the reversed order and therefore, is not effective for $b\geq 1$ systems.

One way to obtain harmonics for $b\geq1$ systems is to change the harmonic expansion in terms of $b^{-1}=\cot(\beta/2)$ \cite{Fairhurst:2019vut}. In this work, we implement an easier solution and that is to simply reverse the harmonic order at $b=1$. Therefore, depending on the $b$ value for a template, we use the following \textit{new} order of the harmonics:
\begin{align}
    \text{for} \hspace{1em} b < 1 &: [h_0,h_1,h_2,h_3,h_4] \\
    b \geq 1 &: [h_4, h_3, h_2, h_1, h_0].
\end{align}

We perform the orthogonalization procedure (equation \ref{Eq:orthogonalization}) only after reversing the harmonics, as this ensures maximum variance is captured on the first step. Once the new orthogonalized harmonics are obtained for templates with $b\geq1$, we perform the matched filtering as previously described. We additionally employ an improved version of the \approximant{IMRPhenomXP} waveform model to reduce the number of harmonics required for approximately $\sim 2.5\%$ of templates that exhibit unphysical features. This issue and the corresponding modifications are discussed in detail in the Appendix \ref{appendix:unphysical-features}.

Figure \ref{fig:Ncomp-comparison} shows the number of harmonics assigned to each template in the mass-ratio $q=m_1/m_2$ and $b$ space for both the original and the updated harmonic prescriptions. With the new scheme, 35,461 templates require fewer harmonics than before, as shown in the pie-charts in Figure \ref{fig:Ncomp-distribution}. Only for systems close to $b\sim1$ multiple harmonics are needed. Reordering the harmonic assignment, we reduce the total number of filters by 65,355, and the number of templates requiring five harmonics drops from 29,580 to just 3. Reversing the harmonics helps massively to improve the efficiency of our search, and the improvement comes in fact with reduced computational costs.  The broader implications of this refinement—particularly its impact on the fitting factors and overall search performance—will be explored in the next section.

We also introduce improvements to the efficiency of the sine-Gaussian $\chi^2$ test. In our precessing search, short-duration noise transients (“blips”) frequently produce triggers that require repeated evaluation of the sine-Gaussian $\chi^2$ veto. To mitigate this, we implemented two key optimizations to the waveform generator. 1) Amplitude-based truncation: we restrict computation to frequency ranges where the waveform amplitude exceeds a precision-based cutoff. By skipping calculations where the exponent is numerically negligible, we significantly reduce the number of operations. 2) Frequency-array caching: we implemented caching for frequency arrays to eliminate redundant memory allocations. The implemented optimizations reduce the computational costs for evaluating sine-Gaussian $\chi^2$ without affecting its discriminatory power.

\begin{figure*}[h]
    \centering
    \includegraphics[width=0.8\linewidth]{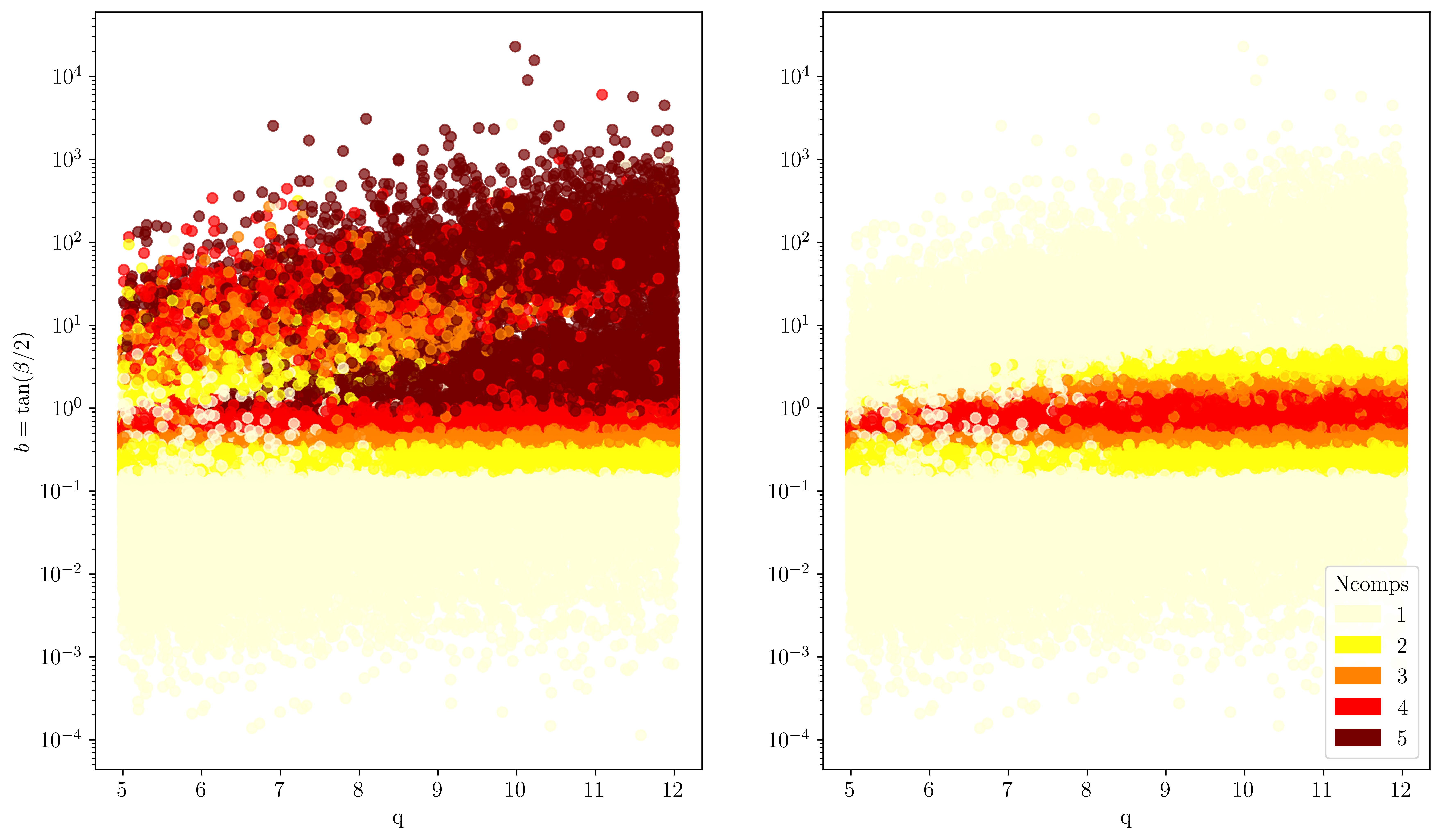}
    \caption{The number of harmonics required by every template in our bank as represented by the $(q, b)$ parameter space, where $q=m_1/m_2$. We show the number of harmonics obtained using the original harmonic order [$h_0,h_1,h_2,h_3,h_4$] (left), and using the new harmonic order (reversed order for $b\geq 1$) (right). The color of each point represents the number of required harmonics: darker templates require more harmonics and lighter the fewer (see legend). Notice the significant reduction in the number of harmonics in the upper region $(b \gtrsim 1 )$ from left to right plot: a total of 35461 templates have fewer harmonics on the with the new harmonic scheme. By incorporating the new harmonic order, the total number of harmonics for the entire bank is reduced by 65355, and the number of templates requiring 5 harmonics reduces from 29580 to only 3. Comparing the total number of filters to the previous precessing BBH search \cite{Schmidt:2024jbp, Schmidt:2024hac}, we have five times fewer filters.}
     \label{fig:Ncomp-comparison}    
\end{figure*}

\begin{figure*}[]
    \centering
    \includegraphics[width=0.7\linewidth]{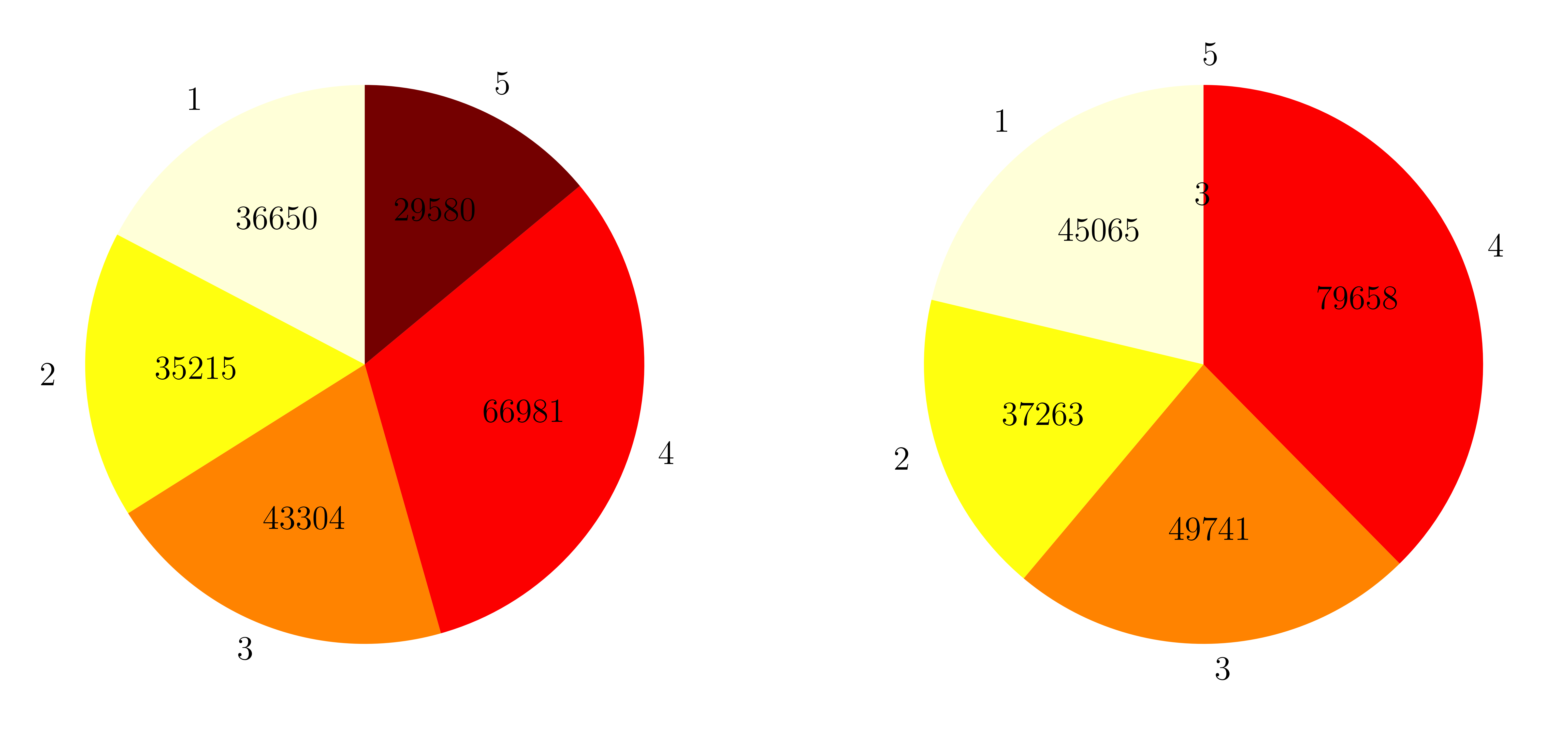}
    \caption{Pie chart of the templates grouped according to their required number of harmonics (denoted by the numbers inside the pie). On the left is the distribution with the original harmonics, and on the right is the distribution with the new harmonic order where we apply the reverse harmonic order for $b \geq1$ templates. Improving to the new scheme, 35461 templates have fewer harmonics. Summing up the numbers in each pie, we find there are in total 65355 additional number of harmonics (filters) with the original harmonic order. With the new scheme the number of templates requiring five harmonics reduces from 29580 to 3.}
     \label{fig:Ncomp-distribution}    
\end{figure*}

\section{Assessing the validity of our search method}

\subsection{Validity of our template bank}
We test the validity of our template bank by checking the fitting factors obtained for a population of simulated binaries. For a faithful comparison of the efficiency of our bank with the bank used in the study \cite{Schmidt:2024jbp}, we use the same distributions to sample the binary parameters. The prior ranges correspond to the same masses and extrinsic parameters as in Table \ref{Table:TB-params} but with different priors on the spin magnitude. In \cite{Schmidt:2024jbp}, nonphysical priors of uniform in the fourth power of $|s|^4$ were used to make sure a large fraction of the population corresponds to highly precessing systems. In addition to the nonphysical prior, we employ the standard prior of uniform in the first power $|s|$ of the spin magnitudes. The prior distributions for both our injected population are described in Table \ref{Table:Injection_set}.

\begin{table}[]
    \centering
    \begin{tabular}{|c|c|c|}
    \hline 
        Parameter & Range & Distribution\\
        \hline
        \hline
        $m_1$ & [15, 70]$M_{\odot}$ & Uniform \\
        \hline
        $m_2$ & [3, 10]$M_{\odot}$ & Uniform \\
        \hline
        $q$ & [5, 12]  & Constraint on the masses\\
        \hline
        Spin-directions & $[0, \pi]$ & Isotropic \\
        \hline
        $\phi_c$ & $[0, 2\pi]$ & Uniform angle \\
        \hline
        $i$ & [$0, \pi$] & Uniform in sin angle\\
        \hline
        $(\alpha, \delta)$ & ($[0, 2\pi]$, $[0, \pi]$) & Uniform sky \\
        \hline
        $\psi$  & $[0, 2\pi]$ & Uniform \\
        \hline
        $d_{\text{L}}^c$  & [30, 350] Mpc & Uniform \\
    \hline
    \end{tabular}\\
    
Highly-precessing
    \begin{tabular}{|c|c|c|}
    \hline
    Parameter & Range & Distribution\\
    \hline
    \hline 
       $|s_1|$ & [0.0, 0.99] & Uniform in $|s_1|^4$\\
        \hline
        $|s_2|$ & [0.0, 0.99] & Uniform in $|s_2|^4$\\
     \hline
    \end{tabular}
    
Normal-precessing
    \begin{tabular}{|c|c|c|}
    \hline 
       $|s_1|$ & [0.0, 0.99] & Uniform in $|s_1|$\\
        \hline
        $|s_2|$ & [0.0, 0.99] & Uniform in $|s_2|$\\
     \hline
    \end{tabular}
    
    \caption{Description of the \textit{highly precessing} injected population of signals (same as the one used in \cite{Schmidt:2024jbp}). We also generate a \textit{normal precessing} injection population that is sampled from the above except the spin magnitudes are distributed uniformly in the first power between the range [0.0, 0.99].}
    \label{Table:Injection_set}
\end{table}

\begin{figure}[]
    \centering
    \includegraphics[width=\linewidth]{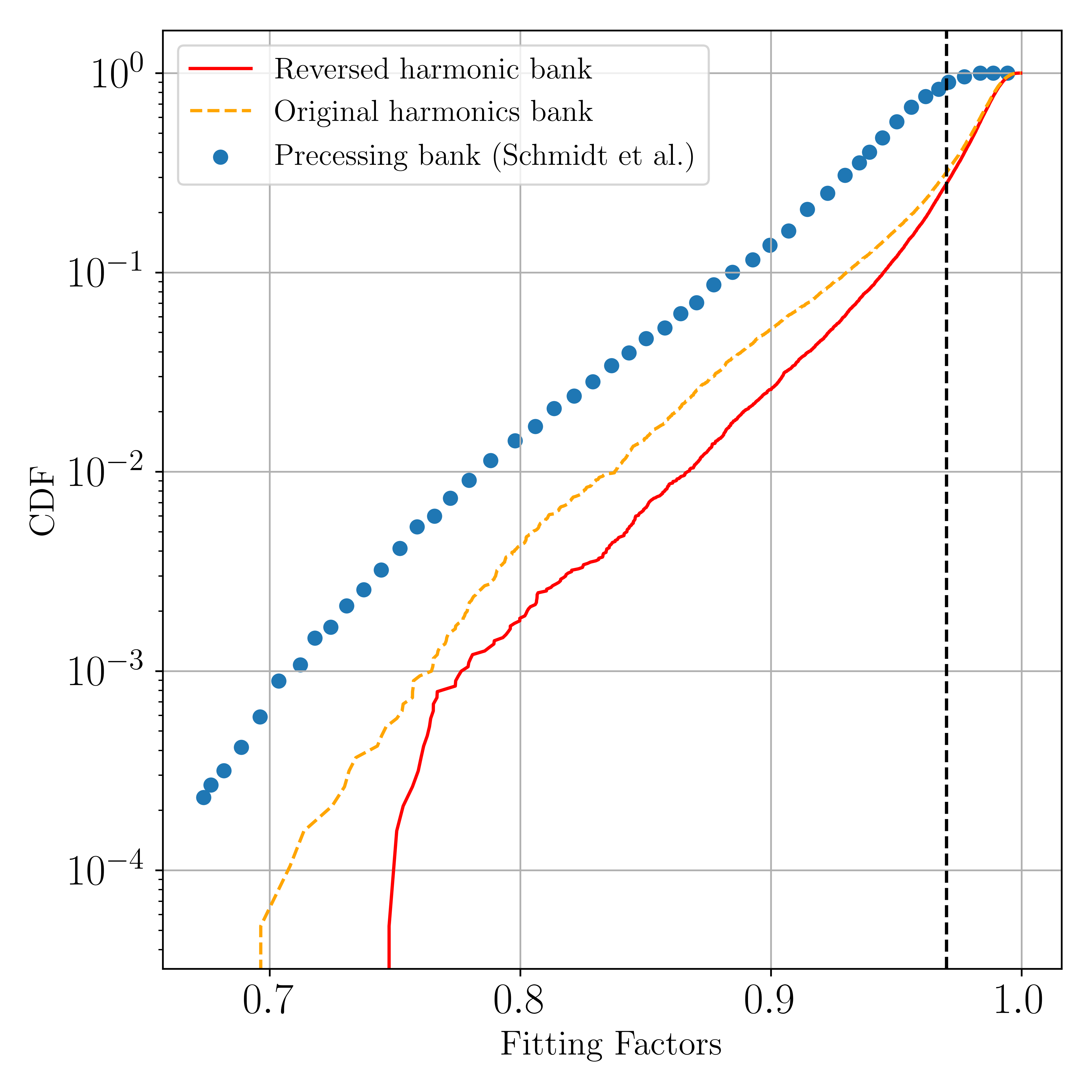}
    \caption{Cumulative distribution of fitting factors for the highly precessing injection set (as described in Table \ref{Table:Injection_set}) obtained by the previous precessing BBH search bank (blue circles) \cite{Schmidt:2024jbp}, by our harmonic bank with original harmonics order (orange dotted line) and the same bank including the new harmonic order(red bold line). The previous precessing search bank contains $\sim 2.3 \times 10^6$ and our bank contains $\sim 2 \times 10^5$ templates. Our bank achieves much better fitting factors with $5 \times$ fewer filters compared to the previous search, after accounting for the extra harmonics for each template (up to three). The black-dotted line corresponds to the minimum match of the bank = 0.97.}
    \label{fig:FF-comparison}    
\end{figure}

\begin{figure}[]
    \centering
    \includegraphics[width=\linewidth]{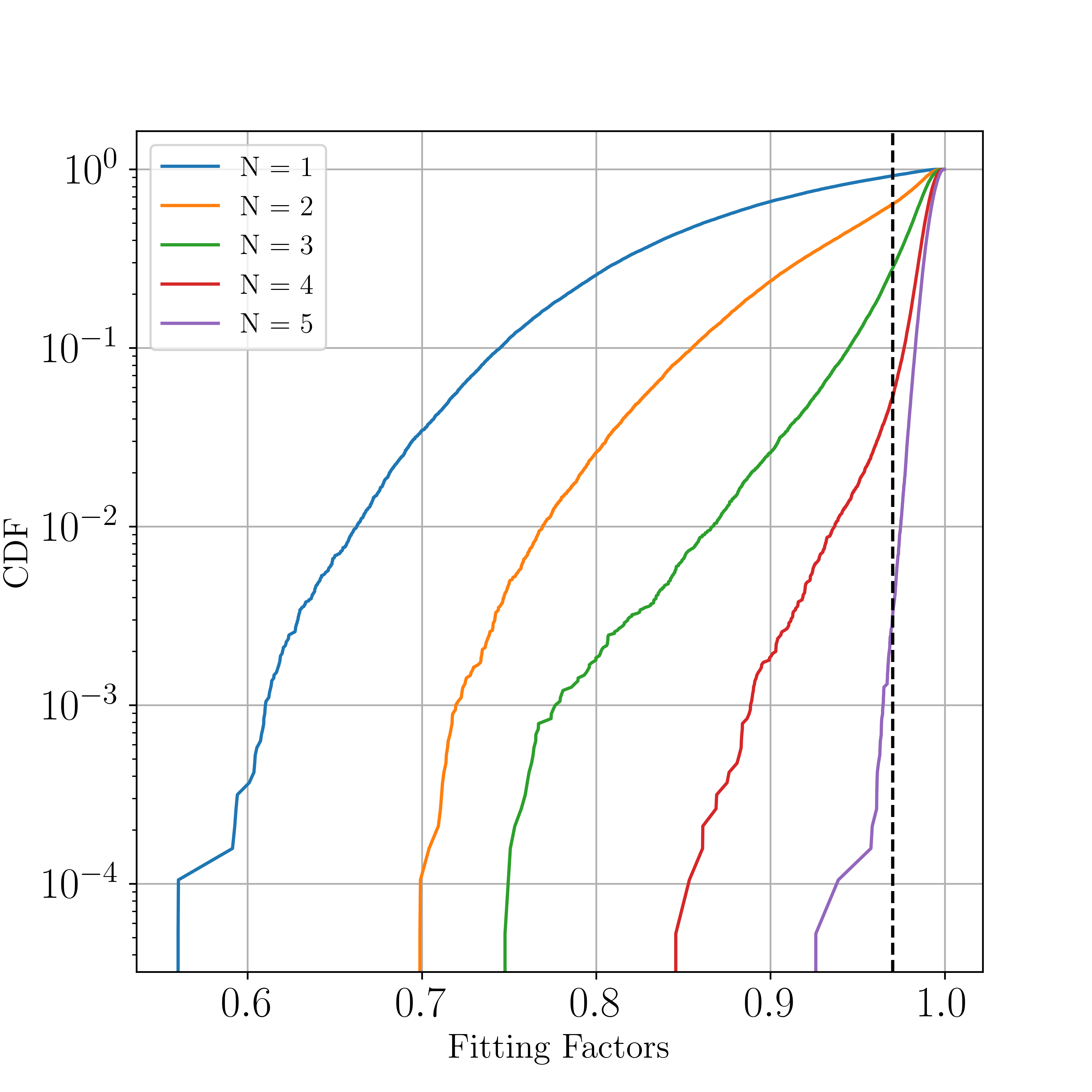}
    \caption{Cumulative distribution of fitting factors for the highly-precessing population of simulated signals (see Table \ref{Table:Injection_set}) obtained using our template bank (with reverse harmonics included). The different curves represent the fitting factor distributions for using the maximum number of harmonics up to the labeled number (legend). Note, the bank becomes more efficient as we employ more harmonics: as evident by decreasing tail of low fitting factors as the number of harmonics increases. In this search we use only up to three harmonics (green curve). The black-dotted line corresponds to the minimum match of the bank = 0.97.}
     \label{fig:FF-nharms}    
\end{figure}

We compute the fitting factors for highly precessing injections using up to three harmonics. In Figure \ref{fig:FF-comparison} we show the cumulative distribution function of the fitting factors obtained via three different banks: 1) bank used in the previous precessing search \cite{Schmidt:2024hac,Schmidt:2024jbp}, 2) our bank with the original harmonic order, 3) our bank including the reverse harmonic order. Our bank with either of the harmonic order produces better fitting factors than the bank in \cite{Schmidt:2024jbp}. With the new harmonic order fitting factors are better than the original harmonics. We also show the fitting factor distribution when choosing different maximum number of harmonics in Figure \ref{fig:FF-nharms}. The fitting factors improve further if including the fourth and the fifth harmonic. However, currently the ranking statistic used in our search is defined only for up to three harmonics.


\input{GWTC-events}

\subsection{Evaluating search sensitivity via Monte-Carlo integration}\label{Sec:MC-sensitivity}
A search pipeline’s efficiency is quantified by its ability to recover a population of simulated signals embedded in the data. This efficiency is expressed in terms of the sensitive space–time volume, $\langle VT \rangle$, which characterizes the fiducial volume within which the pipeline can detect signals over an observation time $T_{\text{obs}}$. The population-averaged sensitive space–time volume is defined as \cite{Mehta:2025jiq}:
\begin{align}
    \langle VT \rangle = T_{\text{obs}} \int \int p(\text{det}| z, \zeta) p(\zeta) \dfrac{dV_c}{dz}\dfrac{1}{1+z}dz d\zeta,
    \label{Eq:expected-detections}
\end{align}
where $dV_c/dz$ is the differential co-moving volume,  $p(\zeta)$ is the astrophysical probability distribution of binary with $\zeta$ parameters (except redshift) and $p(\text{det} | z, \zeta)$ is the probability of detecting the same system situated at redshift $z$ which produces a trigger with an IFAR greater than the specified threshold.  


Signals are injected into the detector data and then recovered using a search pipeline. However, if an astrophysical distribution is used, then a large fraction of injections are not recoverable by any searches since they are placed at distances greater than the reach of current detectors, which leads to poor estimation of the detection probability $p_{\text{inj}}(z, \zeta)$. Therefore, we use a prior that allows better estimation of the detection probability, and then re-weight the above integral to an astrophysical distribution according to the method described in \cite{Tiwari:2017ndi}. 

We sample the chirp-distance $d_L^c = \mathcal{M}_c^{5/6}d_L$ uniformly such that lighter (quieter) systems are placed closer and heavier (louder) systems are placed further away. Also, the mass distributions are chosen uniformly for convenience. The weights are written as the ratio of the probabilities of the astrophysical $p_{\text{astro}}$ and injected $p_{\text{inj}}$ distance (chirp-distance) distributions as:
\begin{align}
    w_i & = \dfrac{p_{\text{astro}}(d^c_{L}[i])}{p_{\text{inj}}(d^c_{L}[i])},\\
    & = p_{\text{astro}}(d_{L}[i])\times (\mathcal{M}_c[i])^{5/6} \\
    & = p_{\text{astro}}(d_{c}[i]) \times (1 + z[i]) \times (\mathcal{M}_c[i])^{5/6},
    \label{Eq:MC-weights}
\end{align}
where $(\mathcal{M}_c)^{5/6}$ and $(1+z)$ factors are the Jacobians for changing the variables from chirp-distance to luminosity distance and then to co-moving distance $d_c$ ($d_L^c \rightarrow d_L \rightarrow d_c$) respectively. For the target astrophysical distribution, we assume the standard prior of uniform in comoving volume which corresponds to $p_{\text{astro}}(d_c) \propto d_c^2$.



Following \cite{Tiwari:2017ndi}, the Monte-Carlo integration of the equation (\ref{Eq:expected-detections}) over the discrete injections is given by 
\begin{align}
    \langle VT \rangle = V_{c}[z^{\text{inj}}_{\text{max}}] \times \dfrac{\sum \limits_{i \in \text{recovered}} \epsilon[\zeta_i]w_i}{\sum \limits_{i \in \text{all}} \epsilon[\zeta_i] w_i}, 
    \label{Eq:VT-MC-sum}
\end{align}
where $\epsilon[\zeta]$ for an injection with $\zeta$ parameters is either 0 (missed) or 1 (recovered) at a given false alarm rate threshold, and $V_c[z^{\text{inj}}_{\text{max}}]$ is the co-moving volume up to the maximum redshift $z_{\text{max}}$ corresponding to the farthest injection in the population
\begin{align}
    V_c[z^{\text{inj}}_{\text{max}}] = \int_{0}^{z_{\text{max}}} \dfrac{dV_c}{dz}\dfrac{1}{1+z}dz.
\end{align}




\section{Search results from O3 data}

We search for the publicly available data from the Advanced LIGO detectors during their third observing run \cite{KAGRA:2023pio}. This observing run was split in two runs O3a lasting six-months from April 1st to October 1st 2019, and O3b with five-months from November 1st 2019 to March 27th 2020. 

We perform matched filtering with a low-frequency cutoff of 30 Hz. We conduct two searches — a precessing search and an aligned-spin search—using the template banks described previously. For each search, we inject two sets of signals, the highly precessing and normally precessing systems, as listed in Table \ref{Table:Injection_set}. We then combine the results from the individual sub-searches with equal weighting to obtain the results for our \textit{combined} search. The list of top candidates, the template parameters associated with each candidate, and the configuration files necessary to reproduce the analysis are available in our data release \cite{github}.

\subsection{Observational results}

In Table \ref{Table:event-list}, we report the list of significant events observed in our search with an inverse FAR greater than 1 per year. We did not find any new events. Our combined search recovered 29 events which were already reported in the previous GWTC catalogs \cite{LIGOScientific:2021usb, KAGRA:2021vkt, LIGOScientific:2025slb}. We miss the GW200129\_065458 event because this event has detector-frame parameters that lie outside our search region.

We also report two sub-threshold (IFAR $<$ 1 [Yr]) events that were not reported in any GWTC catalogs in Table \ref{Table:candidate-list}. 

\begin{table*}[]
  \centering
  \caption{The table lists two candidates from our combined search that have not been previously reported in any GWTC catalogs. For each candidate, we present the parameters from the precessing template yielding the highest ranking statistic. These parameters include the component masses $(m_1, m_2)$, spin projections along the orbital angular momentum $(s_{1z}, s_{2z})$, the effective spin $\chi_\mathrm{eff}$, effective precession parameter $\chi_P$. We also provide the matched-filter signal-to-noise ratios measured in LIGO Hanford $\rho_\mathrm{H}$ and LIGO Livingston $\rho_\mathrm{L}$ .}
    \begin{tabular}{ c  c  c  c  c  c  c  c  c  c  c }
    \hline
     & GPS time & IFAR [yr] & $m_1$ [$M_{\odot}$] & $m_2$[$M_{\odot}$] & $s_{1z}$ & $s_{2z}$ & $\chi_\mathrm{eff}$ & $\chi_P$ & $\rho_\mathrm{H}$ & $\rho_\mathrm{L}$ \\
\hline
1 & 1243736409.19 & 0.365 & 40.82 & 4.31 & 0.64 & 0.02 & 0.58 & 0.60 & 5.71 & 9.02  \\
2 & 1253413448.93 & 0.125 & 35.42 & 3.02 & -0.05 & 0.02 & -0.08 & 0.53 & 6.89	& 7.28  \\
    \hline
    \label{Table:candidate-list}
    \end{tabular}
\end{table*}

\subsection{Search sensitivity}

Aligned-spin searches are always performed, even in highly precessing regions (as the one explored in this work) where they are expected to have reduced sensitivity. Therefore, we explore the efficacy of performing a \textit{combined} search, which is formed by equally combining aligned-spin and precessing search sensitivities. We estimate the improvement in $\langle VT \rangle$ for our combined search and compare it to our aligned-spin only search. We estimate the sensitivity for each search as described by  equation (\ref{Eq:VT-MC-sum}). For an aligned-spin search we use an IFAR threshold of 1 per year or above. Since we are searching over the same data twice in the combined search, we must penalize the significances for each sub-search with a trials factor of two, and therefore, apply an IFAR threshold of 2 [Yr] for each sub-search. This effectively makes the combined search IFAR threshold the same as the aligned-spin only search. The sensitivity ratio is computed as
\begin{align}
    \langle VT \rangle_{\text{ratio}} = \dfrac{\langle VT \rangle^{\text{IFAR} \geq2}_{\text{AS}} + \langle VT \rangle^{\text{IFAR} \geq2}_{\text{Precessing}}}{\langle VT \rangle^{\text{IFAR}\geq1}_{\text{AS}}}
\end{align}
 
\begin{figure*}[]
    \centering
    \includegraphics[width=\linewidth]{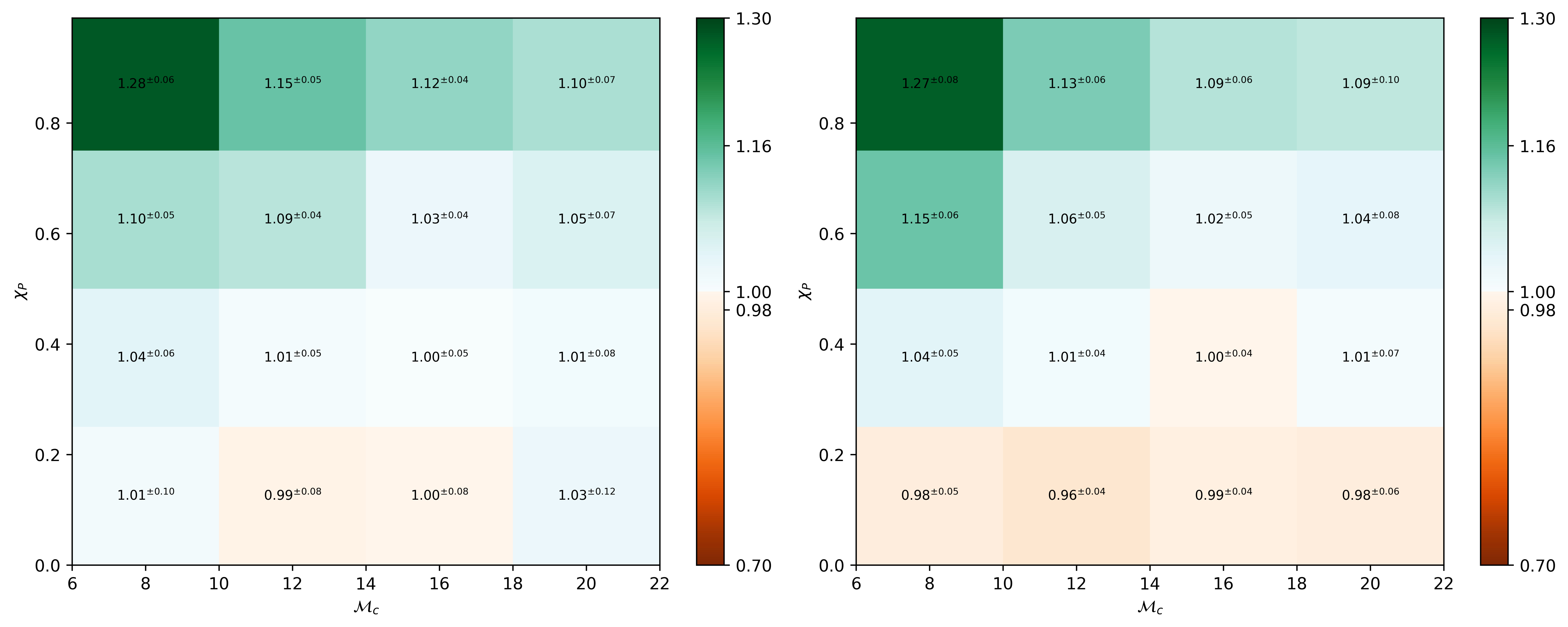}
    \caption{Ratio of the sensitive $\langle VT \rangle$, for a combined (precessing + aligned-spin) search at IFAR threshold of 2.0 [Yr] versus the aligned-spin search at IFAR threshold of 1.0 [Yr] for the \textit{highly precessing} injection set (left) and for \textit{normal precessing} injection set (right). We show the $\langle VT \rangle_{\text{ratio}}$ as a function of ($\mathcal{M}_c, \chi_P$). We find an overall improvement for the parameter space, with the combined search being up to $\sim 28\%$ more sensitive compared to an aligned-spin search only. The combined search becomes effective with increasing precessional effects (increasing $\chi_P$) and, with decreasing chirp mass because precessional effects are better measured for longer duration (i.e. smaller chirp mass) signals.}
     \label{fig:sensitive-VT-ratio}    
\end{figure*}

In Figure \ref{fig:sensitive-VT-ratio}, we show the $\langle VT \rangle_{\text{ratio}}$ for both the 1) highly and 2) normal precessing injection sets (as described in Table \ref{Table:Injection_set}). Our combined search performs consistently better than the aligned-spin–only search. In particular, we observe a gain in sensitivity of up to 28\% for highly precessing systems with $\chi_P \geq 0.5$. This improvement arises because the aligned-spin search loses sensitivity in these regions, whereas the precessing search maintains robust performance. For weakly precessing systems with $\chi_P < 0.2$, the combined search performs on par with the aligned-spin–only search, with only a modest (less than $\sim 4\%$) decrease in sensitivity for aligned-spin injections. The precessing-only search loses up to $\sim 23\%$ sensitivity in low-precessing regions $(\chi_P < 0.2)$ of the search space. Overall, these results demonstrate that our combined strategy delivers a significant enhancement in sensitivity, especially in the regime where precession effects are strong.

\section{Conclusions and future improvements}

In this work, we apply an improved implementation of the harmonic decomposition technique to carry out a full-scale search for precessing binary black holes in the O3 data of the Advanced LIGO detectors. We introduce a novel method for estimating the number of harmonics, which substantially reduces the number of filters required in a harmonic search. Combined with our template bank, this scheme reduces the filter count by a factor of five relative to the previous search for highly precessing binaries \cite{Schmidt:2024hac}, while simultaneously achieving better fitting factors.

We searched over the entire O3 data of Advanced LIGO detectors and found no new significant events. With our precessing search, we have found 29 previously reported events in the latest GWTC-4 catalog \cite{LIGOScientific:2025slb}. Recognizing that aligned-spin searches will continue to be used as standard searches to produce GW catalogs, we quantify the benefit of a joint search that weights precessing and aligned-spin searches equally. Relative to an aligned-spin-only search, our combined search improves the sensitive volume by up to $28\%$ for systems with $\chi_P \geq 0.5$, and yields more modest gains across the rest of the parameter space. This improvement is justified by the observation of events like GW200129\_065458 and GW190521\_074359 which exhibit strong support for $\chi_P > 0.5$ \cite{Islam:2023zzj}.

Aligned-spin searches are highly optimized, no precessing search has developed a ranking statistic that outperforms the performance of aligned-spin searches employed to produce gravitational-wave catalogs. There are two major improvements that can improve the sensitivity of our harmonic search. First, developing a better $\chi^2$ statistic to better distinguish precessing BBH signals from short-duration (blip) glitches. Our current implementation checks for expected power in the signal which increases as the binary approaches merger. However, precessing templates can have modulations such that they have a significant power away from merger or no power at all close to the merger. We have observed in our search, such templates trigger a large number of high SNR false alarms which contaminate our background estimation and reduce the sensitivity of our search. The glitch classes responsible for the increased rate of noise triggers in the precessing search are discussed in Appendix \ref{appendix:glitchy-bkg}. Implementing a novel $\chi^2$ method is therefore crucial for improving precessing searches in the future. 

Secondly, improving the multi-detector ranking statistic. There are multiple components to enhance this statistic a few of which have already been proposed (but not yet implemented) in \cite{McIsaac:2023ijd}. In addition, the phase-time-amplitude histograms required for the ranking statistic, can be extended to include the $b$ parameter to properly account for the uncertainty due to noise. Addition of a new parameter exponentially increases the phase-time-amplitude file sizes, which may pose computational challenges. Novel methods which can easily and accurately learn the distribution of triggers for a population of astrophysical sources for different detector networks would be needed to extend our current implementation. 

Would further enhancing the fitting factors significantly improve the ranking of real precessing events at a given false alarm rate? With the exception of only three templates, the entire template bank satisfies the required number of harmonics. The inclusion of the fourth harmonic leads to a significant improvement in the fitting factors (see Figure \ref{fig:FF-nharms}). However, adding an additional harmonic also increases the degrees of freedom of the matched-filter statistic by two, which in turn leads to a higher false alarm rate. It is therefore worth investigating how the combined search performs when the fourth harmonic is included in a precessing search.


The continued growth of the gravitational-wave event catalog is expected to enhance the prospects of detecting precessing binary systems. Accurate measurement of spin precession will play a crucial role in disentangling the various astrophysical formation channels contributing to the observed population. Given that our combined search has approximately up to $28\%$ greater sensitivity relative to current aligned-spin searches and achieves up to a fivefold improvement in efficiency over existing precessing search methods, it provides a promising framework for future searches for precessing binaries in upcoming gravitational-wave data.

\acknowledgments
The authors would like to thank Charlie Hoy and Laura Nutall for continued discussion throughout this project. We are grateful for the computational resources provided by LIGO Laboratory and the ICG, SEPNet and the University of Portsmouth. This research has made use of data, software and/or web tools obtained from the Gravitational Wave Open Science Center (https://www.gw-openscience.org), a service of LIGO Laboratory, the LIGO Scientific Collaboration, the Virgo Collaboration, and KAGRA. This material is based upon work supported by NSF’s LIGO Laboratory, which is a major facility fully funded by the National Science Foundation. LIGO Laboratory and Advanced LIGO are funded by the United States National Science Foundation (NSF) as well as the Science and Technology Facilities Council (STFC) of the United Kingdom, the Max-Planck-Society (MPS), and the State of Niedersachsen/Germany for support of the construction of Advanced LIGO and construction and operation of the GEO600 detector. The Australian Research Council provided additional support for Advanced LIGO. Virgo is funded through the European Gravitational Observatory (EGO), by the French Centre National de Recherche Scientifique (CNRS), the Italian Instituto Nazionale di Fisica Nucleare (INFN) and the Dutch Nikhef, with contributions by institutions from Belgium, Germany, Greece, Hungary, Ireland, Japan, Monaco, Poland, Portugal, Spain. KAGRA is supported by the Ministry of Education, Culture, Sports, Science and Technology (MEXT), Japan Society for the Promotion of Science (JSPS) in Japan; National Research Foundation (NRF) and the Ministry of Science and ICT (MSIT) in Korea; Academia Sinica (AS) and National Science and Technology Council (NSTC) in Taiwan. For the purpose of open access, the author(s) has applied a Creative Commons Attribution (CC BY) licence to any author accepted manuscript version arising.


\FloatBarrier

\appendix

\section{Unphysical features in \approximant{IMRPhenomXP}}\label{appendix:unphysical-features}

\begin{figure*}[]
    \centering
    \includegraphics[width=0.8\linewidth]{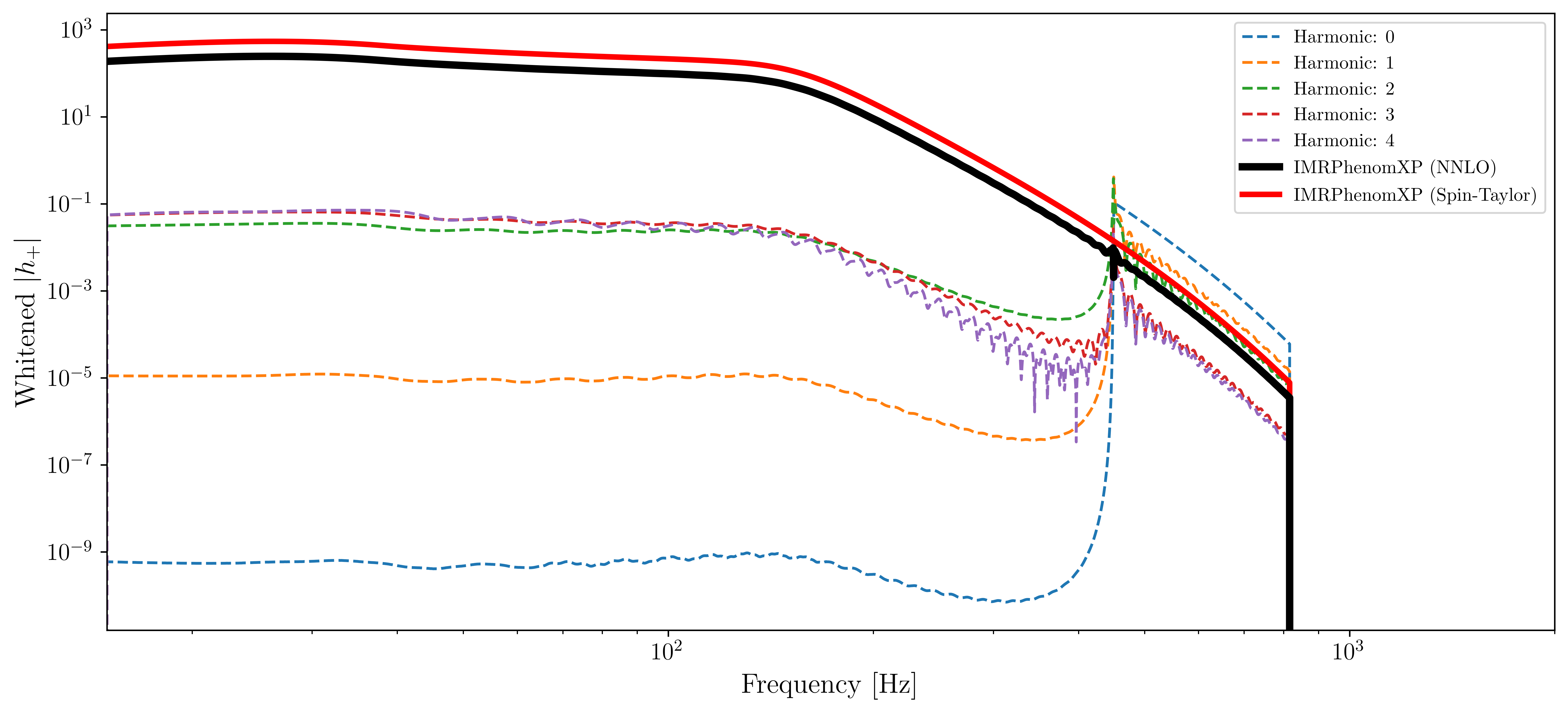}
    \caption{Example of an unphysical feature observed when using the default \approximant{IMRPhenomXP} (NNLO angles) waveform model (black bold line). The waveform corresponds to the system with intrinsic parameters: $(m_1, m_2) = (68.11, 6.462)$, and spins $\vec{s}_1 \simeq [0.0, 0.0, -0.85]$, $\vec{s}_2 \simeq [0.04, 0.02, -0.44]$. The sharp kink observed around $\sim 400$ Hz forces all the harmonics (dashed lines) to increase in amplitude to fit the unphysical feature better -- the number of harmonics required in such cases is always five. When using the more accurate SpinTaylor prescription for the same approximant, the kink disappears (red bold line). The harmonics corresponding to the improved waveform model (not shown) do not peak close to the kink.}
     \label{fig:bad_template}    
\end{figure*}

We find that for 4980 ($\sim 2.5\%$) templates in our bank, the default  prescription of the \approximant{IMRPhenomXP} waveform model shows unphysical sharp kinks in the frequency domain waveforms. An example of this is shown in Figure \ref{fig:bad_template}. The harmonic decomposition of the templates is not an accurate representation of a physical system. In such cases, the harmonics peak in amplitude around the sharp kinks to fit the waveform better, and therefore, all five harmonics are required to reconstruct such unphysical templates. These unphysical features penalize our search in two ways: 1) they lead to more than required number of filters, and 2) the sub-dominant harmonics match with glitches, causing higher rate of false alarms. 

We attribute these kinks to the use of the default prescription for the Euler angles. By default, the next-to-next-to-leading-order (NNLO) prescription is employed to map aligned-spin waveform modes in the co-precessing frame to precessing waveform modes in the inertial frame, as described in \cite{Hannam:2013oca}. However, previous studies have shown that the NNLO prescription fails to accurately capture complex precessional dynamics, such as transitional precession \cite{Yu:2023lml}. We find that this limitation affects a significant portion of our search parameter space, leading to unphysical features in some templates. To address this issue, we adopt the more accurate, albeit slightly slower, SpinTaylor prescription for the Euler angles \cite{Colleoni:2024knd}, which successfully removes these unphysical features (see also Figure \ref{fig:bad_template}). The SpinTaylor implementation of \approximant{IMRPhenomXP} is at most a factor of four slower than the default NNLO version. We use the improved waveform model to generate the full template bank and the filters used for matched filtering. 


\section{Increased rate of background triggers}\label{appendix:glitchy-bkg}

\begin{figure*}[]
    \centering
    \includegraphics[width=\linewidth]{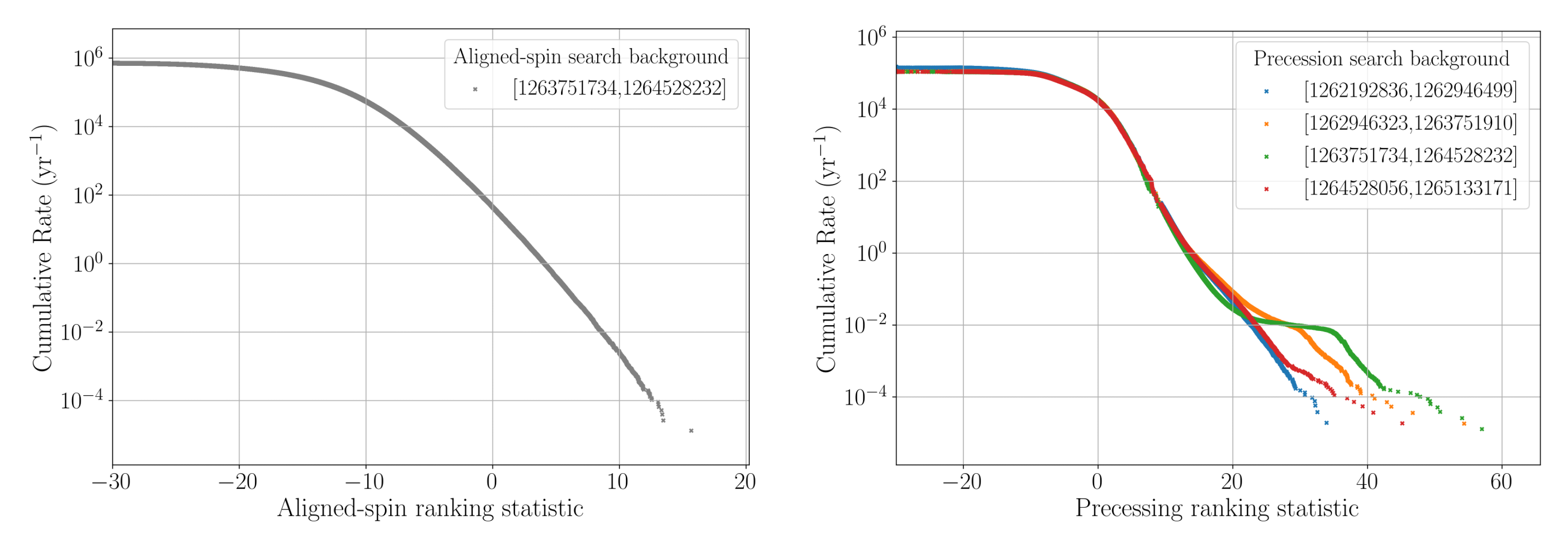}
    \caption{Background distributions of coincident (HL) ranking statistics for noise triggers in the aligned-spin (left) and precessing (right) searches. The vertical axis shows the false-alarm rate, while the horizontal axes show the coincident ranking statistics used in each search. Because the two searches use different ranking statistics, we do not make a direct quantitative comparison between them. Qualitatively, the precessing search exhibits a higher rate of noise triggers. This behavior arises from the increased degrees of freedom in the precessing single-detector ranking statistic, which raises the effective noise floor. For the aligned-spin search, the cumulative background—shown for GPS times [1263751734, 1264528232]—falls off exponentially with increasing ranking statistic and remains consistent throughout the O3 run. In contrast, the precessing search shows extended background tails during periods of increased glitch rates. We show the background distribution over four data chunks. Note the indicated kinks in the green and orange curves. These kinks arise from glitches such as blips and scattered-light noise that the current implementation of the $\chi^2$ tests does not effectively down-rank, leading to an excess of high-ranking noise triggers and an increased false-alarm rate.}
     \label{fig:background-dists}    
\end{figure*}

Our precessing search has a higher rate of noise (background) triggers than for the aligned-spin search. The additional degrees of freedom in the single-detector ranking statistic raise the noise floor, leading to a huge number of noise triggers surviving the vetoing and coincidence steps. As a result, the precessing search is more susceptible to producing loud coincident noise events than the aligned-spin search as shown in Figure \ref{fig:background-dists}. The coincident ranking statistics for the two searches are different, which prevents a direct quantitative comparison along the horizontal axes, however, a clear qualitative distinction is observed in the vertical axes and the tails of the distributions. The precessing search shows a pronounced excess of high-ranking noise triggers, corresponding to an elevated false-alarm rate at large statistic values.

\begin{figure*}[]
    \centering
    \includegraphics[width=\linewidth]{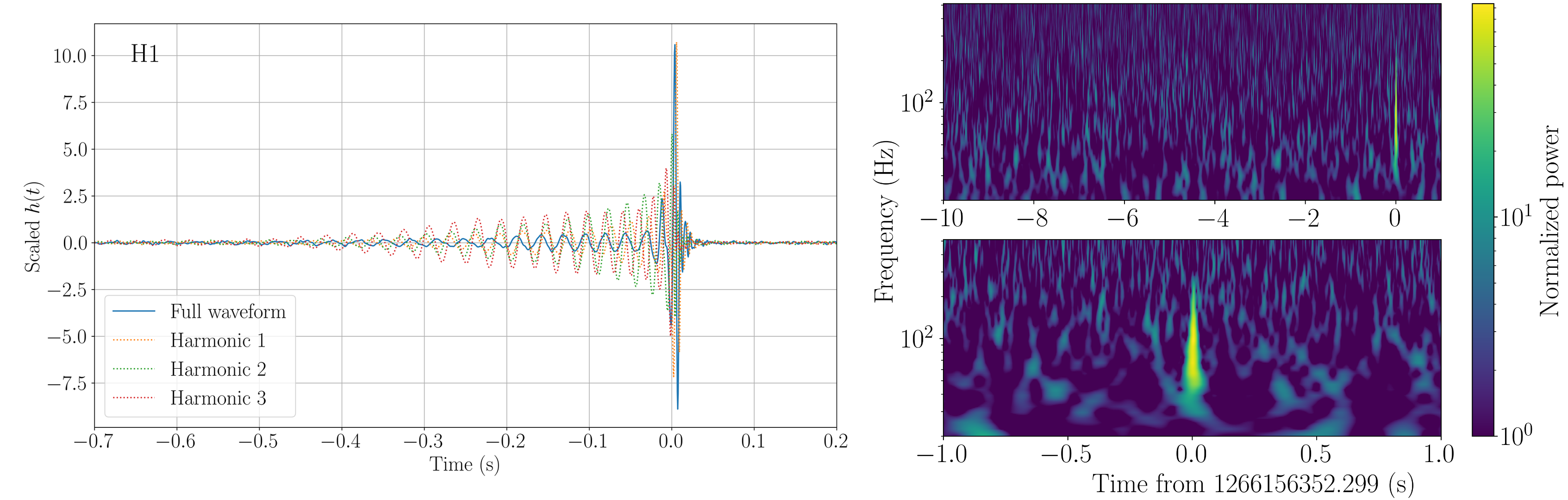}
    \caption{Example of an extremely short duration template triggering a blip glitch in LIGO Hanford. The template corresponds to highly-asymmetrical  $(m_1^{\text{det}}, m_1^{\text{det}}) = (59, 5)M_{\odot}$ source. Significant power in the signal is only during the last $\sim0.01$ second which matches the blip glitch as visible in the spectrograms (second column). Since the template is extremely short duration, our current chi-squared tests struggle to penalize such single detector triggers.}
     \label{fig:blip-glitch}    
\end{figure*}

\begin{figure*}[]
    \centering
    \includegraphics[width=\linewidth]{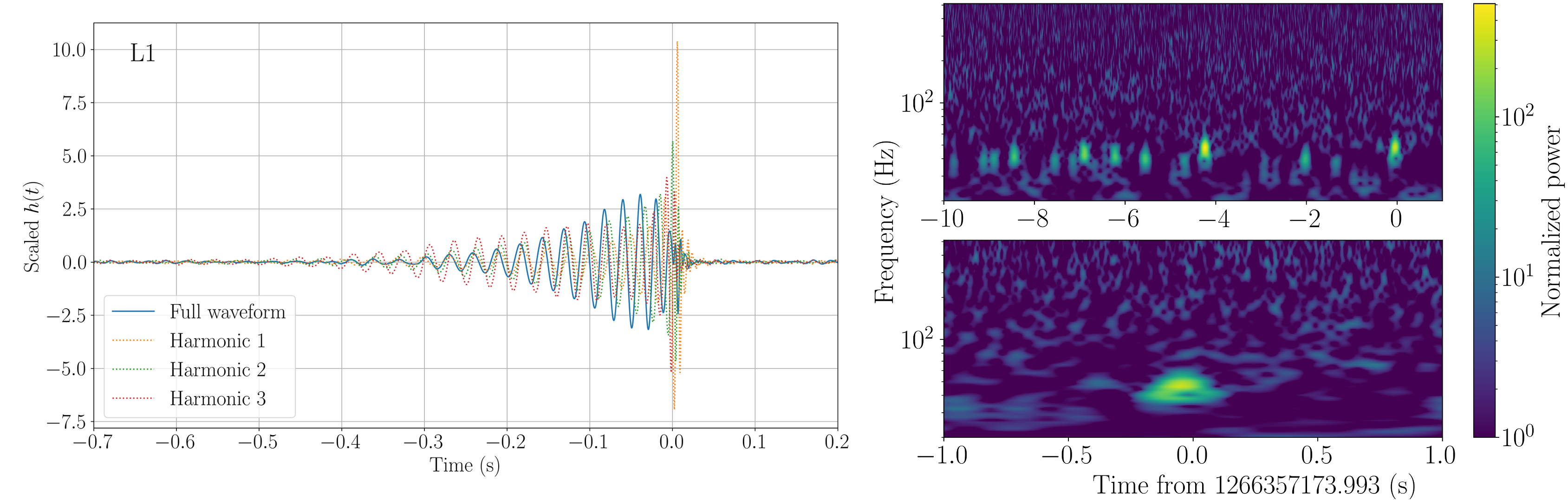}
    \caption{Example of a template that exhibits significant power away from the merger. This template is identical to the one shown in Figure \ref{fig:blip-glitch} and corresponds to a highly asymmetric source with detector-frame masses $(m_1^{\text{det}}, m_2^{\text{det}}) = (59, 5),M_{\odot}$. Such templates typically match low-frequency scattering noise. Because the current implementation of the $\chi^2$ test is designed to identify increasing power with increasing gravitational-wave frequency, peaking near merger, the search has limited ability to effectively down-rank triggers of this type.}
     \label{fig:scattering-glitch}    
\end{figure*}

A further contribution to the elevated false-alarm rate in the precessing search arises from the limited effectiveness of the current signal-consistency (chi-squared) tests in suppressing certain classes of instrumental glitches. We find there are two classes of glitches that severely impact the sensitivity of our search: 1) \textit{blip} glitches 2) \textit{scattering} glitches. Short-duration templates, such as those corresponding to highly asymmetric, high-mass and/or highly-precessing systems, can closely match blip glitches, with significant signal power as shown in Figure \ref{fig:blip-glitch}. In Figures \ref{fig:blip-glitch} and \ref{fig:scattering-glitch}, we show template corresponding to a highly asymmetrical and highly spinning source that matched with different class of glitches leading to a coincident trigger that contaminated our background. Due to the extremely short duration ( power confined to the final $\sim 0.01$ s of the waveform) of the template in Hanford, the chi-squared test struggle to sufficiently penalize such triggers in individual detectors. We also find, templates with substantial power away from the merger can couple strongly to low-frequency scattering noise (shown in Figure \ref{fig:scattering-glitch}). Since the present chi-squared implementation primarily tests for increasing power with increasing gravitational-wave frequency, it is not optimally designed to down-rank triggers dominated by pre-merger or low-frequency power. Together, these effects lead to an accumulation of high-ranking noise triggers and highlight key limitations of the current veto strategy in the presence of instrumental artifacts.

\FloatBarrier

\bibliography{references}
\end{document}

%% file: GWTC-events.tex
\begin{table*}[]
  \centering
  \caption{This table lists the 29 GWTC events recovered by our combined search with an inverse FAR greater than 1 per year. For each recovered event, we present the GPS time, the IFAR [Yr] obtained by the PyCBC-BBH analysis in the GWTC-2.1 \cite{LIGOScientific:2021usb} or GWTC-3 \cite{KAGRA:2021vkt} catalogs, and the IFAR [Yr] obtained by our combined search and the precessing only search. In the combined search, we assign equal weight to the aligned-spin and precessing-only searches; consequently, the reported IFAR corresponds to the mean of the two sub-search IFARs.}
    \begin{tabular}{ c  c  c  c  c}
    \hline
    Event & GPS time & GWTC PyCBC IFAR & Our combined search IFAR & Precessing only search IFAR\\
    &  & [Yr] & [Yr] & [Yr]\\
\hline
GW190408\_181802 & 1238782700.28 & $8.3\times10^3$ & $2.7\times 10^4$ & $1.3 \times 10^4$\\
GW190412\_053044 & 1239082262.17 & $8.3\times10^3$ & $8.0\times 10^3$ & $1.6 \times 10^4$\\
GW190503\_185404 & 1240944862.29 & 384.6 & 10.0 & 20 \\
GW190512\_180714 & 1241719652.42 & $9.1 \times 10^3$ & $4.4 \times 10^3$ & 140\\
GW190519\_153544 & 1242315362.40 & $9.0\times10^3$ & 4.1 & 7.2\\
GW190521\_030229 & 1242442967.46 & 769.2 & 1.8 & 3.6\\
GW190521\_074359 & 1242459857.46 & $4.4\times10^4$ & $5.6\times 10^4$ & $5.6 \times 10^4$\\
GW190706\_222641 & 1246487219.34 & 2.9 & 96.5 & 180\\
GW190707\_093326 & 1246527224.17 & $5.3\times10^4$ & $7.6\times 10^4$ & $5.1 \times 10^4$\\
GW190720\_000836 & 1247616534.70 & $1.3\times10^4$ & $1.2\times 10^4$ & 11.0\\
GW190725\_174728 & 1248112066.46 & 0.3 & 3.3 & 1.6\\
GW190727\_060333 & 1248242631.98 & $5\times10^3$ & 4.1 & 8.1\\
GW190728\_064510 & 1248331528.53 & $1.3\times10^4$ & $3.2\times 10^4$ & $4.0 \times 10^3$\\
GW190828\_063405 & 1251009263.75 & $1.4\times10^4$ & $3.6\times 10^4$ & $4.8 \times 10^3$\\
GW190828\_065509 & 1251010527.88 & $9.0\times10^3$ & 510.5 & 21.0\\
GW190915\_235702 & 1252627040.69 & $1.4\times10^4$ & $6.0\times 10^2$ & 100\\
GW190924\_021846 & 1253326744.84 & $1.2\times10^4$ & 600 & $5.6 \times 10^3$ \\
GW190930\_133541 & 1253885759.24 & 83.3 & 20.5 & 27\\
GW191105\_143521 & 1256999739.93 & 27.8 & 103.8 & 7.6\\
GW191109\_010717 & 1257296855.22 & 21.3 & 7.2 & 2.4\\
GW191129\_134029 & 1259070047.20 & $4.2\times10^4$ & $3.8\times 10^4$ & $3.8 \times 10^4$\\
GW191204\_171526 & 1259514944.09 & $8.3\times10^4$ & $7.0\times 10^4$ & $6.9 \times 10^4$\\
GW191215\_223052 & 1260484270.33 & 3.6 & 126.9 & 3.8\\
GW191222\_033537 & 1261020955.12 & $1.0\times10^4$ & 4.2 & 8.3\\
GW200128\_022011 & 1264213229.89 & 232.5 & 1.1 & 2.2\\
GW200224\_222234 & 1266618172.40 & $1.3\times10^4$ & $4.8\times 10^4$ & $8.6 \times 10^3$\\
GW200225\_060421 & 1266645879.39 & $2.4\times10^4$ & $1.5\times 10^4$ & 45\\
GW200311\_115853 & 1267963151.39 & $1.3\times10^4$ & $3.7\times 10^4$ & $6.8 \times 10^3$\\
GW200316\_215756 & 1268431094.16 & 1.72 & 48.0 & 2\\
    \hline
    \label{Table:event-list}
    \end{tabular}
\end{table*}